\documentclass[]{aa}  

\usepackage{natbib}
\bibpunct{(}{)}{;}{a}{}{,}
\usepackage{graphicx}
\usepackage{revsymb}	
\usepackage{amsmath}	
\usepackage[usenames]{color}	
\usepackage{epstopdf}	
\usepackage{txfonts}
\usepackage{amssymb}
\usepackage{color}
\usepackage[justification=centering]{caption}
\usepackage{geometry} 
\usepackage[parfill]{parskip} 
\usepackage{amssymb}
\usepackage{slantsc}
\usepackage{array}
\usepackage{url}
\usepackage{amsfonts}
\usepackage[colorlinks=true,linkcolor=blue, urlcolor=blue, citecolor=blue]{hyperref}
\usepackage{sidecap}
\usepackage{graphicx}
\usepackage{xcolor}
\usepackage{nicefrac}
\usepackage{subfigure}
\usepackage{epsfig}
\colorlet{rouge}{red!70!darkgray}

\begin{document}
\title{Higher metal abundances do not solve the solar problem}
\author{G. Buldgen\inst{1,2}\and P. Eggenberger\inst{1}\and A. Noels\inst{2} \and R. Scuflaire\inst{2} \and A. M. Amarsi\inst{3} \and N. Grevesse \inst{2,4} \and S. Salmon \inst{1}}
\institute{D\'epartement d'Astronomie, Universit\'e de Gen\`eve, Chemin Pegasi 51, CH-1290 Versoix, Switzerland \and STAR Institute, Université de Liège, Liège, Belgium \and Theoretical Astrophysics, Department of Physics and Astronomy, Uppsala University, Box 516, 751 20 Uppsala, Sweden \and Centre Spatial de Liège, Université de Liège, Angleur-Liège, Belgium }
\date{October, 2022}
\abstract{The Sun acts as a cornerstone of stellar physics. Thanks to spectroscopic, helioseismic and neutrino flux observations, we can use the Sun as a laboratory of fundamental physics in extreme conditions. The conclusions we draw are then used to inform and calibrate evolutionary models of all other stars in the Universe. However, solar models are in tension with helioseismic constraints. The debate on the ``solar problem'' has hitherto led to numerous publications discussing potential issues with solar models and abundances.}
{Using the recently suggested high-metallicity abundances for the Sun, we investigate whether standard solar models, as well as models with macroscopic transport reproducing the solar surface lithium abundances and analyze their properties in terms of helioseismic and neutrino flux observations.}
{We compute solar evolutionary models and combine spectroscopic and helioseismic constraints as well as neutrino fluxes to investigate the impact of macroscopic transport on these measurements.}
{When high-metallicity solar models are calibrated to reproduce the measured solar lithium depletion, tensions arise with respect to helioseismology and neutrino fluxes. This is yet another demonstration that the solar problem is also linked to the physical prescriptions of solar evolutionary models and not to chemical composition alone.}{A revision of the physical ingredients of solar models is needed in order to improve our understanding of stellar structure and evolution. The solar problem is not limited to the photospheric abundances if the depletion of light elements is considered. In addition, tighter constraints on the solar beryllium abundance will play a key role in the improvement of solar models.}
\maketitle
\section{Introduction}

The Sun plays a key role in stellar physics. Thanks to the numerous high-quality observations available, its acts as both a laboratory of fundamental physics and a calibrator for stellar evolution models \citep{JCD2021}. However, the modelling of solar structure is still a subject of debate, fueled in part by uncertainties in the solar chemical composition. Striking disagreements exist with helioseismic constraints, when adopting the ``low-metallicity'' compositions presented in \citet{AGS2005} and \citet{AGSS09},  and more recently in \citet{Asplund2021} and \citet{Amarsi2021} --- hereafter, the ``solar problem'' \citep[see e.g.][and references therein]{Basu08,Buldgen2019R}. These low-metallicty compositions are based on spectroscopic analyses of the solar disc-centre intensity, utilising three-dimensional radiative-hydrodynamic simulations of the solar atmosphere (3D RHD) and where available non-local thermodynamic equilibrium radiative transfer (non-LTE).

Recently, \citet{Magg2022} presented a new spectroscopic analysis of the solar chemical composition. In contrast to the papers above, their analysis was based on the solar disc-integrated flux (in which, spectral lines form higher up in the atmosphere, and thus are potentially more sensitive to non-LTE effects  as well as to blends that can be exacerbated by the extra broadening due to rotation), and 1D model atmospheres (derived from spatial and temporal averages of 3D RHD models). They inferred a ``high-metallicity'' chemical composition similar to the canonical 1D LTE compilations of \citet{GrevNoels} and \citet{GS1998}, hereafter GN93 and GS98. Using this high-metallicity composition and standard solar models, they found better agreement with helioseismic constraints. They concluded that the solar problem is solved, without need for any revision of fundamental physical ingredients. 

We consider whether the solar problem is in fact solved. We show that a revision of abundances simply affects the magnitude of the corrections required in solar models, but does not validate a physical prescription for modelling the Sun. Recently, \citet{Eggenberger2022} demonstrated that stepping away from the standard solar models is required to simultaneously reproduce both helioseismic inversions of the solar internal rotation and the spectroscopic measurement of the lithium photospheric abundance; a result that was foreseen by \citet{JCD1996}. In the last decades, numerous studies have investigated the implications of revising the physical ingredients of solar models, such as accretion, mass-loss, transport of chemicals, convection, opacities, dark matter, dark energy and nuclear reactions \citep[see e.g.][and refs therein]{Guzik2001,Brun2002,Guzik2005,Guzik2010,Vinyoles2015,Spada2018,Zhang2019,Yang2019,
Bellinger2022,Saltas2022,Yang2022} and multiple generations of standard and non-standard models were computed \citep[e.g.][amongst other]{Serenelli2010,Vinyoles2017, JCD2018,Jorgensen2018}. 

We focus on the changes induced by reproducing the lithium abundance with various parametric diffusion coefficients using solar models built with the \citet{Magg2022} abundances, similarly to \citet{Richard1996}. We discuss the impact on neutrino fluxes and helioseismic constraints of reproducing all available spectroscopic constraints and how improved beryllium abundances will help us pin down the properties of macroscopic transport at the base of the solar convective zone. 

\section{Standard and non-standard solar models}\label{Sec:Models}

We present our set of solar evolutionary models and discuss the changes of properties induced by the inclusion of macroscopic transport at the base of the solar convective zone (BCZ). We used both Standard Solar Models (SSM) and non-standard models, computed with the Liège Stellar Evolution Code \citep{ScuflaireCles}. We used the following physical ingredients: the solar abundances were the \citet{Magg2022} solar abundances, the opacity were the OP opacities \citep{Badnell2005}, supplemented at low temperature by opacities of \citet{Ferguson}, the FreeEOS equation of state was adopted \citep{Irwin}, and the  nuclear reaction rates were from \citet{Adelberger}. 

The first model, ``Model Std'', is a SSM, including microscopic diffusion following \citet{Thoul}, with the screening coefficients of \citet{Paquette} and the effects of partial ionization. As seen from Table \ref{tabModels}, the results for this setup are almost identical to those illustrated in Table 6 of \citet{Magg2022}. The slight differences in the positioning of the BCZ are due to differences in the prescription for microscopic diffusion that can alter the metallicity profile close to the BCZ (see also Table 1 in \citet{Buldgen2019} for an illustration).

All other models in Table \ref{tabModels} were computed with an additional parametric diffusion coefficient, aiming at reproducing the solar lithium abundance \citep{Wang2021}. To do so, we assumed that the initial lithium and beryllium abundances were provided by the meteoritic values \citep{Lodders2009}. We start with Model $DT_{R}$ that is fitted to reproduce the transport induced by the combined effect of shear-induced turbulence, meridional circulation and the magnetic Tayler instability \citep{Eggenberger2022}. We adopted a simple parametrization in power law of the density following \citet{Proffitt1991}
\begin{align}
D_{T}(r)=D\left(\frac{\rho_{BCZ}}{\rho(r)}\right)^{n},
\end{align}
with $D$ a constant in $cm^{2}s^{-1}$, $\rho$ the local value of the density, $\rho_{BCZ}$ the value of the density at the BCZ of the model and $n$ a fixed constant number. In their recent paper, \citet{Eggenberger2022} found that the behaviour of rotating models could be well reproduced with $n=1.3$, which is used in Model $DT_{R}$. In addition, we also tested different values of $n$, from $2$ to $5$ to investigate the impact of reproducing the solar lithium depletion in solar models. We note that the depletion is highly significant, as lithium is reduced by $2.29$ dex with an uncertainty on the current photospheric abundance of $0.06$ dex. Therefore it constitutes an important constraint to consider when studying solar evolution. 

The results of the calibrated models with parametric transport are provided in Table \ref{tabModels}. The BCZ position in the model is altered when macroscopic mixing is included, thus we investigated in model $DT_{R}+Ov$ the effect of including adiabatic overshooting to replace the transition in temperature gradient at the helioseismic value of $0.713\pm0.001R_{\odot}$ \citep{Basu97BCZ}.
\begin{table}[h]
\caption{Parameters of the solar evolutionary models}
\label{tabModels}
  \centering
%\resizebox{\linewidth}{!}{
\begin{tabular}{r | c | c | c }
\hline \hline
\textbf{Name}&\textbf{$\left(r/R\right)_{\rm{BCZ}}$}&\textbf{$\left( m/M \right)_{\rm{CZ}}$}&\textbf{$\mathit{Y}_{\rm{CZ}}$} \\ \hline
Model Std&$0.7148$&$0.9760$& $0.2443$ \\
Model $DT_{R}$&$0.7182$&$0.9769$& $0.2522$\\ 
Model $DT_{2}$&$0.7178$&$0.9768$& $0.2513$ \\
Model $DT_{3}$&$0.7177$&$0.9768$& $0.2510$ \\
Model $DT_{4}$&$0.7176$&$0.9768$& $0.2507$ \\
Model $DT_{5}$&$0.7174$&$0.9767$& $0.2503$ \\
Model $DT_{R}+Ov$&$0.7133$&$0.9759$& $0.2516$ \\
\hline
\end{tabular}
%}
\end{table}

%\begin{table*}[t]
%\caption{Parameters of the solar evolutionary models}
%\label{tabModels}
%  \centering
%\begin{tabular}{r | c | c | c | c | c | c | c | c }
%\hline \hline
%\textbf{Name}&\textbf{$\left(r/R\right)_{\rm{BCZ}}$}&\textbf{$\left( m/M \right)_{\rm{CZ}}$}&\textbf{$\mathit{Y}_{\rm{CZ}}$}&\textbf{$\mathrm{m}_{0.75}$} & $\phi(\rm{pp})$& $\phi(\rm{Be})$&$\phi(\rm{B})$&$\phi(\rm{CNO})$\\ \hline
%Model Std&$0.7148$&$0.9760$& $0.2443$ & $0.9826$ & $5.96$ & $4.89$ & $5.42$ &$6.11$\\
%Model $DT_{R}$&$0.7182$&$0.9769$& $0.2522$& $0.9827$ & $5.98$ & $4.73$ & $5.04$ & $5.53$\\ 
%Model $DT_{2}$&$0.7178$&$0.9768$& $0.2513$ & $0.9827$ & $5.98$ & $4.74$ & $5.07$ & $5.57$\\
%Model $DT_{3}$&$0.7177$&$0.9768$& $0.2510$ & $0.9827$ & $5.98$ & $4.75$ & $5.08$ & $5.59$\\
%Model $DT_{4}$&$0.7176$&$0.9768$& $0.2507$ & $0.9827$ & $5.98$ & $4.75$ & $5.09$ & $5.61$\\
%Model $DT_{5}$&$0.7174$&$0.9767$& $0.2503$ & $0.9827$ & $5.98$ & $4.76$ & $5.10$ & $5.63$ \\
%Model $DT_{R}+Ov$&$0.7133$&$0.9759$& $0.2516$ & $0.9827$ & $5.98$ & $4.74$ & $5.06$ & $5.56$\\
%\hline
%\end{tabular}
%\end{table*}

The following sections will discuss the results obtained and their consequences for the current issues in solar modelling. 

\section{Lithium, helium, neutrinos and convective envelope position}\label{Sec:LiHeNeut}

\begin{figure*}
	\centering
		\includegraphics[width=17cm]{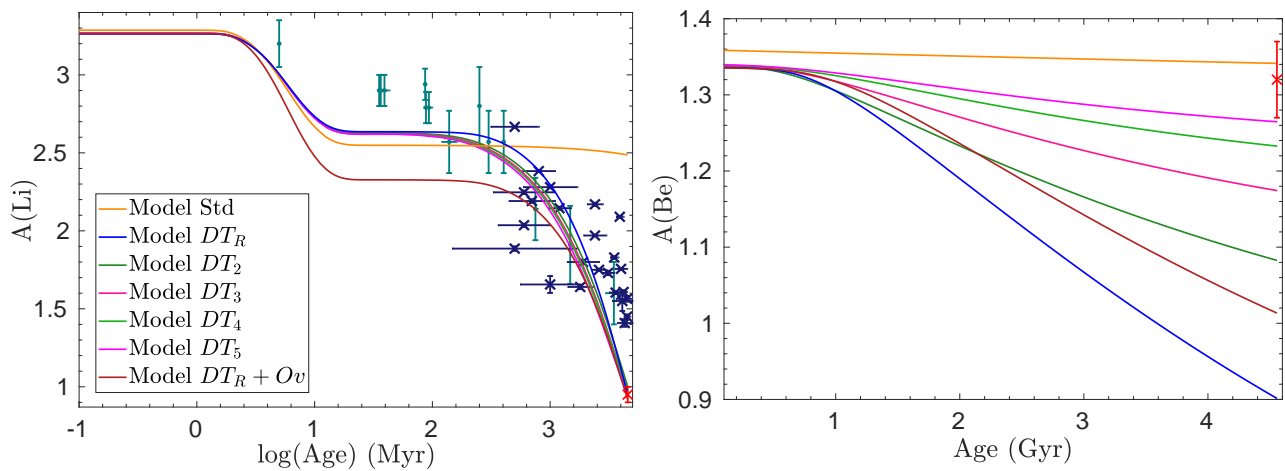}
	\caption{Left panel: lithium photosperic abundance as a function of age for the models of Table \ref{tabModels}, the red cross is the solar value and the dark blue and teal crosses are the values for young solar-like stars in open clusters from \citet{Dumont2021}. Right panel: beryllium photospheric abundance as a function of age for the models of Table \ref{tabModels}, the red cross indicates the solar value from \citet{AGSS09}.}
		\label{Fig:LiBe}
\end{figure*}

The evolution of the photospheric lithium and beryllium abundances as a function of the solar age are shown in Fig. \ref{Fig:LiBe}, with $A(X)=\log(X/H)+12$. A first result confirmed here is that SSMs are unable to reproduce the lithium depletion in the Sun. As mentioned in \citet{Proffitt1991}, additional mixing at the BCZ is required to reproduce the observed depletion. 

The calibration of this mixing was done for various values of $n$, changing the value of $D$ simutaneously to reproduce the lithium abundance. Each leads to a different beryllium depletion at the age of the Sun. In the right Panel of Fig. 1, we show that a higher value of $n$ leads to a lower depletion of beryllium at the age of the Sun. This is a direct consequence of the higher burning temperature of beryllium at $\approx 3.5 \times 10 ^{6}$ K. A higher $n$ value leads to a steeper diffusion coefficient and thus a less efficient transport of beryllium down to $\approx 3.5 \times 10 ^{6}$ K, despite the recalibration of the factor $D$ to reproduce the lithium depletion. Thus beryllium acts as a strong constraint on the functional form of the macroscopic transport coefficient at the BCZ and is thus of highest importance to constrain the physical origin of the lithium depletion. 

The final beryllium abundance will also be affected by the presence or absence of strong adiabatic overshooting at the BCZ. The inclusion of this additional mixing has strong consequences for solar models. First, the position of the BCZ is significantly shifted by about $0.002$ $R_{\odot}$ (hence $2\sigma$) with respect to the position obtained in the SSM framework. This shift in the position of the BCZ is also linked to a small change in the mass coordinate of the convective zone. It is actually due to a change in the metallicity profile close to the BCZ. When only microscopic diffusion is included in the models, the competition between pressure diffusion and thermal diffusion leads to a drop in diffusion velocities close to the BCZ that induces an accumulation of elements \citep[see][for a discussion]{Baturin2006}. 

\begin{figure*}
	\centering
		\includegraphics[width=17cm]{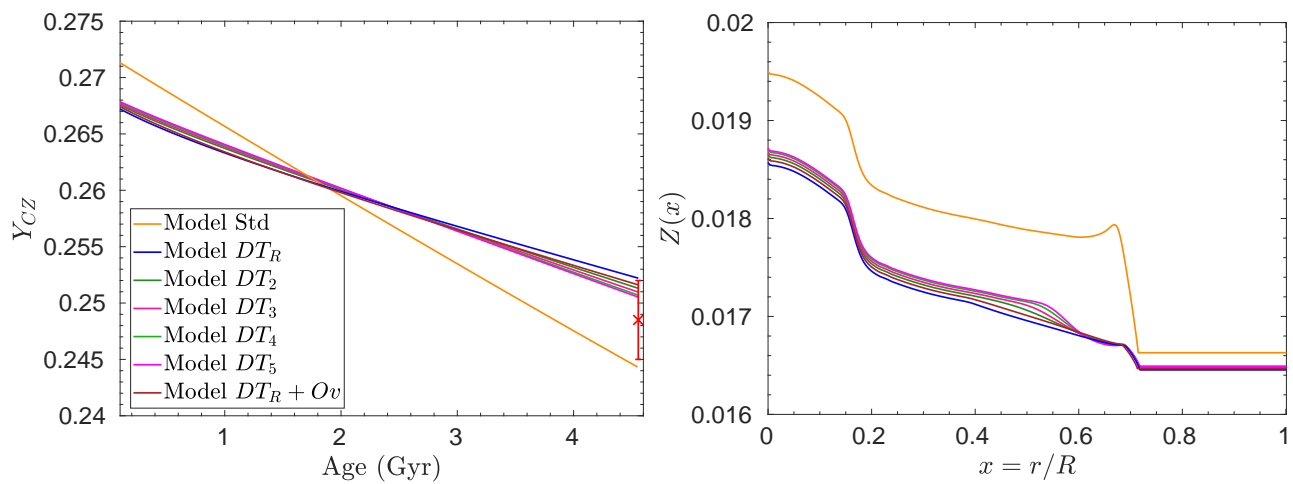}
	\caption{Left panel: evolution of the Helium abundance in the convectine zone $Y_{CZ}$ for the evolutionary models of Table \ref{tabModels}, the red cross indicating the seismic value and the $1\sigma$ interval from \citet{BasuYSun}. Right panel: metallicity profile as a function of normalized radius for the evolutionary models of Table \ref{tabModels}.}
		\label{Fig:YZ}
\end{figure*}

This is particularly visible for the Z profile of SSMs. The accumulation of metals close to the BCZ locally increases the opacity, as the major contributors at the BCZ are Oxygen, Iron and Neon. Therefore, the temperature gradient is locally steepened, leading to a deeper convective zone. Once macroscopic transport is included in the models, this local maximum is erased, which leads to a shallower convective zone. This result is obtained for all models including transport with a $n \leq 5$. From Fig \ref{Fig:YZ}, we can see that $n$ should be much higher than $5$ to avoid this issue, meaning that $D$ should be increased significantly to compensate and the transport should probably behave almost instantaneously in a shallow region, coming closer to the behaviour of the model with overshooting and showing tension with the results of young solar-like stars in open clusters \citep[taken from][]{Dumont2021}. Meanwhile, beryllium would provide a definitive answer on the behaviour of the mixing and thus its physical origin. 

A second consequence of the inclusion of macroscopic mixing is the reduction of the efficiency of gravitational settling. As the microscopic diffusion velocities drop very fast in the radiative interior, if mixing is included, settling is inhibited, as seen in Fig. \ref{Fig:YZ}. To reproduce the solar luminosity and radius at the solar age, a higher initial hydrogen abundance is required and thus the core metallicity at the age of the Sun is reduced. 

Due to the lower core metallicity, neutrino fluxes are significantly affected, as shown in Table \ref{tabNeutrinos}. The pp flux, $\phi_{pp}$, is unchanged as it is mostly related to reproducing the solar luminosity. The beryllium and boron neutrino fluxes, $\phi_{Be}$ and $\phi_{B}$, are significantly affected by the inclusion of macroscopic transport, as the core metallicity, temperature and temperature gradient are not high enough and steep enough to reproduce the observations. We refer the reader to \citet{Salmon2021} for an in-depth discussion in the case of SSM using various physical ingredients as well as to the seminal works by \citet{Bahcall2005b,Bahcall2005c} and to \citet{Villante2021} for a review. Similarly, the CNO neutrino flux is significantly reduced and now in disagreement with the observed value from Borexino \citep{Appel2022}. Therefore, additional processes such as planetary formation \citep{Kunitomo2021, Kunitomo2022} or modification to key physical ingredients such as opacity or the electronic screening formalism \citep{Mussack2011A,Mussack2011B} might be required to reproduce the neutrino observations when the lithium depletion is reproduced, particularly the fluxes from the Borexino experiment. 

\begin{table}[h]
\caption{Neutrino fluxes of the evolutionary models}
\label{tabNeutrinos}
  \centering
\resizebox{\linewidth}{!}{
\begin{tabular}{r | c | c | c | c }
\hline \hline
\textbf{Name}& $\phi(\rm{pp})$& $\phi(\rm{Be})$&$\phi(\rm{B})$&$\phi(\rm{CNO})$\\ \hline
Model Std& $5.96$ & $4.89$ & $5.42$ &$6.11$\\
Model $DT_{R}$& $5.98$ & $4.73$ & $5.04$ & $5.53$\\ 
Model $DT_{2}$& $5.98$ & $4.74$ & $5.07$ & $5.57$\\
Model $DT_{3}$& $5.98$ & $4.75$ & $5.08$ & $5.59$\\
Model $DT_{4}$& $5.98$ & $4.75$ & $5.09$ & $5.61$\\
Model $DT_{5}$ & $5.98$ & $4.76$ & $5.10$ & $5.63$ \\
Model $DT_{R}+Ov$& $5.98$ & $4.74$ & $5.06$ & $5.56$\\
O-G21$^{1}$ & $5.97^{+0.0037}_{-0.0033}$ & $4.80^{+0.24}_{-0.22}$ & $5.16^{+0.13}_{-0.09}$ & $-$\\
Borexino$^{2}$ & $6.1^{+0.6}_{-0.7}$ & $4.99^{+0.13}_{-0.14}$ & $5.68^{+0.39}_{-0.41}$ & $6.6^{+2.0}_{-0.9}$\\
\hline
\end{tabular}}
\small{\textit{Note:} $^{1}$ \citet{OrebiGann2021} , $^{2}$ \citet{Borexino2018}, \citet{Borexino2020}, \citet{Appel2022}}
\end{table}

Another consequence of the inhibition of settling seen from the left panel of Fig \ref{Fig:YZ} is that the mass fraction value of Helium in the CZ is significantly changed. While it was in marginal disagreement with the lower end of the helioseismically inferred interval in the SSM, it is now in marginal agreement or disagreement with the upper end of the interval in the models with transport. This impacts the agreement with the first adiabatic exponent profile, $\Gamma_{1}=\frac{\partial \ln P}{\partial \ln \rho}\vert_{S}$, in the solar convective envelope determined from helioseismology. As seen from \citet{Vorontsov13} (e.g. Fig 6 and 7), a Helium mass fraction above $0.25$ is never in agreement with a metal mass fraction above $0.012$, whatever the equation of state used. Further investigations of the properties of the $\Gamma_{1}$ profile in the solar envelope with the most recent equations of state are required to restrict the $Y-Z$ interval allowed in solar models, as it would provide strong constraints on the transport of chemical elements during solar evolution. 

\section{Helioseismic inversions}\label{Sec:Inversions}

\begin{figure*}
	\centering
		\includegraphics[width=17cm]{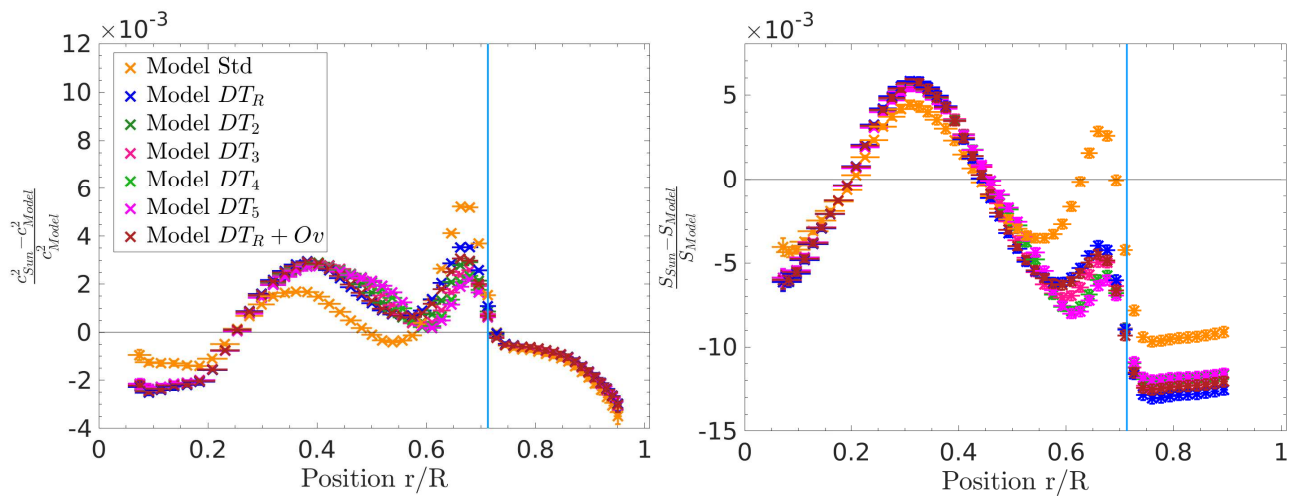}
	\caption{Left panel: relative differences in squared adiabatic sound speed between the Sun and the evolutionary models of Table \ref{tabModels}. Right panel: relative differences in entropy proxy between the Sun and the evolutionary models of Table \ref{tabModels}.}
		\label{Fig:c2SInv}
\end{figure*}

The second point to investigate is the impact of transport on helioseismic inversions. As shown in \citet{Magg2022}, the increased metallicity brought back the models to the level of agreement of the SSMs of the 1990s. As shown in the left panel of Fig \ref{Fig:c2SInv}, we reach similar conclusions. In the right panel of Fig \ref{Fig:c2SInv}, we illustrate the inversion result for the entropy proxy, $S_{5/3}=P/\rho^{5/3}$, introduced in \citet{BuldgenS}, which provides a complementary view on solar models. The level of agreement of the standard model is excellent and similar to that of the GS98 or GN93 models in \citet{Buldgen2019}. This is no surprise as the \citet{Magg2022} abundances are almost the same as the GS98 abundances. They are however in strong disagreement with the surface lithium abundance.

The models based on high-metallicity abundances that are able to reproduce the surface lithium abundance draw a more complex picture. As illustrated in previous works \citep[e.g.][]{Brun02,JCDOV,JCD18}, macroscopic transport reduces the glitch at the BCZ, but also leads to increased discrepancies in the deeper radiative zone and in the core. While still small, these remain significant. From the entropy proxy inversions, we see that macroscopic transport does not improve the agreement with helioseismic inversions. The entropy plateau in the convective zone is now too high, and some deviations are seen in the radiative zone, particularly at the BCZ due to the less steep temperature gradient induced by the erasement of the metallicity peak resulting from microscopic diffusion. This further emphasizes the tension between the models including macroscopic transport and constraints such as neutrino fluxes and helioseismic inversions. 

Overall, the results simply illustrates the importance of putting helioseismic inversions in their context. The final sound speed and entropy proxy profiles will be the result of the whole calibration procedure and therefore of all the ingredients entering the model computation. 

The only way to directly constrain the solar metallicity using helioseismology is by using $\Gamma_{1}$, which is highly sensitive to the equation of state of the solar plasma and has a tendency to favour a low metallicity \citep{Vorontsov13,BuldgenZ} value in the most recent studies. Therefore, determining solely the solar metallicity from helioseismic data is currently restricted by the data available (further analyses using the sets of modes by \citet{Reiter2020} are required) and by the reliability of the solar equation of state in the deep convective envelope. 

\section{Conclusion}

In this study, we discussed in details the implications of the physical ingredients of solar models regarding the conclusion of the solar problem in light of the high-metallicity composition presented by \citet{Magg2022}. We have shown that, as it stands, further constraints on the solar beryllium abundance are key to better understand the properties of macroscopic transport at the base of the solar convective zone. We have also shown that standard solar models are in complete disagreement with the observed solar lithium abundance. However, our work only tackles one aspect of the issue, following previous works \citep[e.g.][]{Richard1996,Brun2002,JCD2018,Zhang2019}, and numerous other processes can be demonstrated to impact the results of comparisons between solar models and helioseismic constraints, as shown in the extensive litterature on the subject \citep[see][and refs therein]{JCD2021}.

One solution to erase this disagreement is to introduce macroscopic transport at the base of the envelope, which could originate in the combined actions of rotation and magnetic instabilities, as shown in \citet{Eggenberger2022}. We tested various parametrizations and their impact on various key constraints of solar models. We find the following points:
\begin{itemize} 
\item The inclusion of macroscopic transport leads to tension with the helioseismic $Y_{CZ}$ value, as well as with the $\Gamma_{1}$ profile inversions of \citet{Vorontsov13}. Looking at the results of Table 1 in \citet{Buldgen2019}, using other key ingredients such as the OPAL opacities or the SAHA-S equation of state might lead to a higher $Y$ value that would further increase these tensions. 
\item The $r_{CZ}$ value is significantly impacted by the inclusion of macroscopic transport to reproduce the photospheric lithium abundance, leading to a shift by about $2\sigma$ of its position. We also note that it is significantly affected by the physical prescription used for microscopic diffusion. 
\item Some tension is also observed with the lithium abundance of young solar twins in stellar clusters. Repositioning $r_{CZ}$ at the helioseismic value using adiabatic overshooting further increases this tension. 
\item By changing the initial conditions of the calibration, macroscopic transport also leads to a significantly lower $Z$ abundance in the core, leading to tension with the CNO Borexino neutrino fluxes and disagreement at $1\sigma$ of the beryllium and boron fluxes.  
\item Sound speed and entropy proxy profiles in the solar interior are overall worse when macroscopic transport is included.  
\end{itemize} 
In their work, \citet{Magg2022} conclude that \textit{"While SSMs offer an incomplete description of the physics in the solar interior, current results alleviate the need for more complex physics, such as accretion of metal-poor material \citep{Serenelli2011}, energy transport by dark matter particles \citep{Vincent2015}, revision of opacities \citep{Bailey}, enhanced gravitational settling and other effects \citep{Guzik2010}."}. We find this assessment to be incorrect, as reproducing the lithium abundance with various parametrizations essentially reduces and even destroys the agreement they find for some constraints. 

The situation for solar models using the \cite{Asplund2021} abundances is going to be slightly different. Including transport helps reproducing the helium abundance in the CZ \citep{Yang2019, Yang2022, Eggenberger2022} and agrees with $\Gamma_{1}$ inferences by \citet{Vorontsov13}. However, it does not solve the issues with the positioning of the BCZ, the sound speed profile and the neutrino fluxes. Those will likely require revision of physical ingredients. Ultimately, the only difference between the \cite{Asplund2021} models and the \cite{Magg2022} models is the required magnitude of these revisions. 

Therefore, we find that the need for additional physics such as the effects of rotation, of which we have a clear map \citep[e.g.][]{SchouRota}, planetary formation, of which the impact can be measured \citep{Kunitomo2021,Kunitomo2022} and opacity revision, which can be quantified and tested \citep[e.g.][]{Ayukov2017,Buldgen2019} is not alleviated. On the contrary, the discussion related to the metal content of the Sun cannot be separated from other fundamental physical ingredients entering solar models such as the equation of state, the radiative opacities, the transport of chemicals or the evolutionary history. In this context, further refining fundamental physics computations, experimental setups and helioseismic techniques are paramount to provide the most accurate description of the internal structure of the Sun and use it as a stepping stone for the modelling of solar-like stars. Indeed, the need for additional transport at the base of convective envelopes is found in solar twins \citep{Deal2015} and F-type stars \citep{Verma2019} and often considered a key ingredient to determine reliable stellar ages. Further exploring such processes for our very own Sun thus appears a natural and meaningful endeavour.

\section*{Acknowledgements}

G.B. is funded by the SNF AMBIZIONE grant No 185805 (Seismic inversions and modelling of transport processes in stars). P.E. and S. J. A. J. S. have received funding from the European Research Council (ERC) under the European Union's Horizon 2020 research and innovation programme (grant agreement No 833925, project STAREX). A.M.A. gratefully acknowledges support from the Swedish Research Council (VR 2020-03940). We acknowledge support by the ISSI team ``Probing the core of the Sun and the stars'' (ID 423) led by Thierry Appourchaux. 

\bibliography{biblioarticleMagg}

\end{document}